\documentclass[a4paper]{jpconf}
\usepackage{graphicx}
\begin{document}
\title{Effects of quark chemical equilibration on thermal photon elliptic flow}

\author{Akihiko Monnai}

\address{RIKEN BNL Research Center, Brookhaven National Laboratory, Upton, NY 11973, USA}

\ead{amonnai@riken.jp}

\begin{abstract}
Large hadronic elliptic flow $v_2$ is considered as an evidence for the existence of a strongly-coupled QGP fluid in high-energy heavy-ion collisions. On the other hand, direct photon $v_2$ has recently been found to be much larger than hydrodynamic estimations, which is recognized as ``photon $v_2$ puzzle". In this study, I discuss the implication of late production of quarks in an initially gluon-rich medium because photons are coupled to quarks. Numerical analyses imply that thermal photon $v_2$ can be visibly enhanced. This indicates that interplay of equilibration processes and collective expansion would be important.
\end{abstract}

\section{Introduction}
Azimuthal anisotropy of particle spectra is a unique observable in heavy-ion collisions. The second harmonics of Fourier expansion of hadronic spectra, elliptic flow $v_2$, is proposed as a useful quantity in non-central collisions because it would reflect the magnitude of interaction in the quark-gluon medium \cite{Ollitrault:1992bk,Poskanzer:1998yz}. Experimental data of Au-Au collisions at $\sqrt{s_{NN}} = 200$ GeV at RHIC show sizable hadronic $v_2$ which is in remarkable agreement with fluid dynamic description, leading to an understanding that the bulk medium is strongly-coupled in the vicinity of QCD crossover. This is reconfirmed in Pb-Pb collisions at $\sqrt{s_{NN}} = 2.76$ TeV at LHC, and higher-order harmonics $v_n$, which comes from geometrical fluctuation, is also shown to be consistent with hydrodynamic description \cite{Schenke:2010rr}.

\begin{figure}[tbh]
\includegraphics[width=16pc]{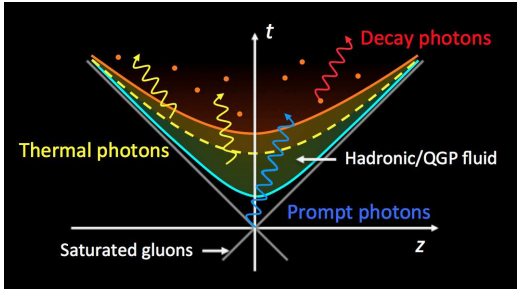}\hspace{2pc}%
\begin{minipage}[b]{16pc}\caption{\label{label}A schematic picture of the high-energy heavy-ion collision from the viewpoint of photons. Decay photons originate from hadronic decay. \label{fig:1}}
\end{minipage}
\end{figure}

Given the success of hydrodynamic models so far, excess of direct photon $v_2$ compared with the model predictions was a surprising discovery \cite{Adare:2011zr,Lohner:2012ct}. Here direct photons are defined as the sum of thermal photons, which are emitted softly from the medium, and prompt photons, which are created in the initial hard processes (Fig.~\ref{fig:1}). The QCD medium is considered to be transparent in terms of electromagnetic interaction. Direct photon $v_2$ was expected to be much smaller than hadronic one because the momentum anisotropy of photons would be acquired only indirectly through that of emission sources, \textit{i.e.}, quarks in the QGP phase and hadrons in the hadronic phase. There also is an experimental indication that photon $v_3$ is large. Some of the possible approaches to the problem would be categorized as follows: (i) modified thermal photon emission mechanism, (ii) modified prompt photon emission mechanism, (iii) other sources of photon emission, (iv) unknown medium-photon interaction, (v) modified bulk evolution and (vi) improvement of experimental data. It should be noted that a combination is also possible. Here I focus on (i) and (v) and discuss the effects of quark chemical equilibration in the bulk medium on thermal photon emission rate in the early stages of hydrodynamic evolution because color glass theory indicates that the medium can be initially gluon-rich \cite{McLerran:1993ni,McLerran:1993ka}.

\section{Hydrodynamics in chemical non-equilibrium}

Most modern hydrodynamic models assume that thermal and chemical equilibrations occur simultaneously even though the precise mechanism of local equilibration is not well known. Since the colliding nuclei are consider to be saturated gluons, there are indications that chemical equilibration can be slower than thermalization \cite{Monnai:2014xya}, a requirement for defining the temperature in hydrodynamics. If the quark number is smaller than that at equilibrium at early times, photon emission is suppressed because photons are coupled to quarks but not to gluons. This could be an enhancement mechanism for thermal photon $v_2$ because the contribution of late photons with larger anisotropy is effectively enhanced \cite{Monnai:2014kqa,Monnai:2014dya}.

The chemically non-equilibrated hydrodynamic model is formulated as follows.
The equation of motion for the system is energy-momentum conservation $\partial_\mu T^{\mu \nu} = 0$ and rate equations for the quark and the gluon number currents
\begin{eqnarray}
\partial_\mu N_q^\mu &=& 2 r_b n_g - 2 r_b \frac{n_g^\mathrm{eq}}{(n_q^\mathrm{eq})^2} n_q^2 , \label{eq:dnqdt} \\
\partial_\mu N_g^\mu &=& (r_a - r_b) n_g - r_a \frac{1}{n_g^\mathrm{eq}} n_g^2 + r_b \frac{n_g^\mathrm{eq}}{(n_q^\mathrm{eq})^2} n_q^2 + r_c n_q - r_c \frac{1}{n_g^\mathrm{eq}} n_q n_g \label{eq:dngdt},
\end{eqnarray} 
where the processes $g \rightleftharpoons g + g$, $g \rightleftharpoons q+\bar{q}$ and $q(\bar{q}) \rightleftharpoons q(\bar{q})+g$ (denoted by the subscripts $a$, $b$ and $c$, respectively) are considered. Note that the four-gluon vertex process is suppressed by additional $\alpha_s$. Here the net baryon density is assumed to be vanishing. An inviscid case is considered for simplicity, though introduction of viscosity to the formalism is straight-forward. $n_q$ and $n_g$ are the quark and the gluon number densities defined in the tensor decomposition $N^\mu_{q} = n_q u^\mu$ and $N^\mu_{g} = n_g u^\mu$ where $u^\mu$ is the flow. $r_a$, $r_b$ and $r_c$ are the chemical reaction rates. Since the pair production/annihilation is the only quark number changing process considered here, $1/r_b$ roughly corresponds to the quark chemical relaxation time of the system. $n_q^\mathrm{eq}$ and $n_g^\mathrm{eq}$ are the equilibrium densities introduced for the parton numbers to be conserved in equilibrium. They are determined using the parton gas picture with $N_f = 2$. Thermalization is assumed to be quick enough to restore thermal equilibrium right after the splitting and merging processes occur. Note that the above rate equations are valid only in the QGP phase. The system is simply assumed to be equilibrated below the crossover temperature.

QGP photon emission rate is modified by simply factoring the phase-space distributions by $n_q/n_q^\mathrm{eq}$ or $n_g/n_g^\mathrm{eq}$ according to Ref.~\cite{Traxler:1995kx}. Hadronic photon emission rate \cite{Turbide:2003si,Arleo:2004gn} is unchanged because the system is assumed to be in chemical equilibrium in the hadronic phase. They are smoothly connected by a hyperbolic function at $T_c = 0.17$ GeV with the width $\Delta T_c = 0.017$ GeV \cite{Monnai:2014kqa}. 
Boost-invariant (2+1)-dimensional ideal hydrodynamics is considered for the background medium. Hydrodynamic input for numerical estimations is summarized in Table~\ref{table:1}. 
\begin{table}[h]
\small
\caption{\label{tc}Input to chemically non-equilibrated hydrodynamic model \label{table:1}} 
\centering
\begin{center} 
\begin{tabular}{ll} 
\br 
Fluid properties&Description\\ 
\mr
Hadronic EoS & Hadron resonance gas ($m_h \leq 2.5$ GeV)\\  
QGP EoS & Parton gas ($N_f = 2$)\\
Chemical reaction rate & $r_i = c_i T$ ($i=a,b,c$)\\
\br 
Initial conditions&Description\\ 
\mr
Gluon energy density $e_g(\tau_0,x,y)$& Glauber model (Au-Au at $\sqrt{s_{NN}} = 200$ GeV, $b = 7$ fm) \cite{Kolb:2000sd}\\  
Quark energy density $e_q(\tau_0,x,y)$& 0 GeV$\cdot$fm$^{-3}$\\  
Gluon number density $n_g(\tau_0,x,y)$& $n_g^\mathrm{eq} + n_q^\mathrm{eq}/2$\\  
Quark number density $n_q(\tau_0,x,y)$& 0 fm$^{-3}$\\  
Initial time $\tau_0$& 0.4 fm/$c$\\
\br 
\end{tabular} 
\end{center}
\end{table}

\section{Numerical results}

\begin{figure}[tbh]
\includegraphics[width=13.5pc]{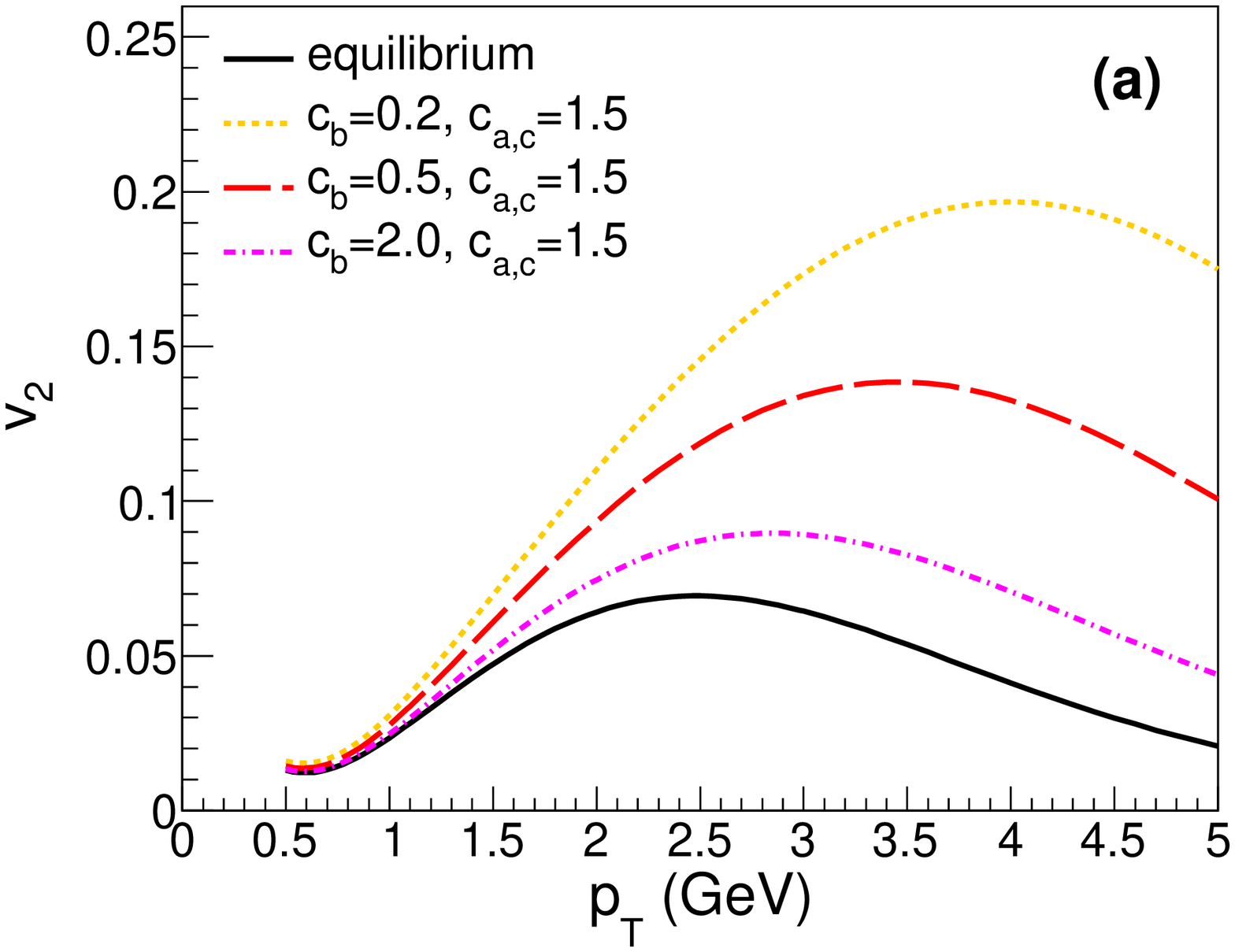} 
\includegraphics[width=13.5pc]{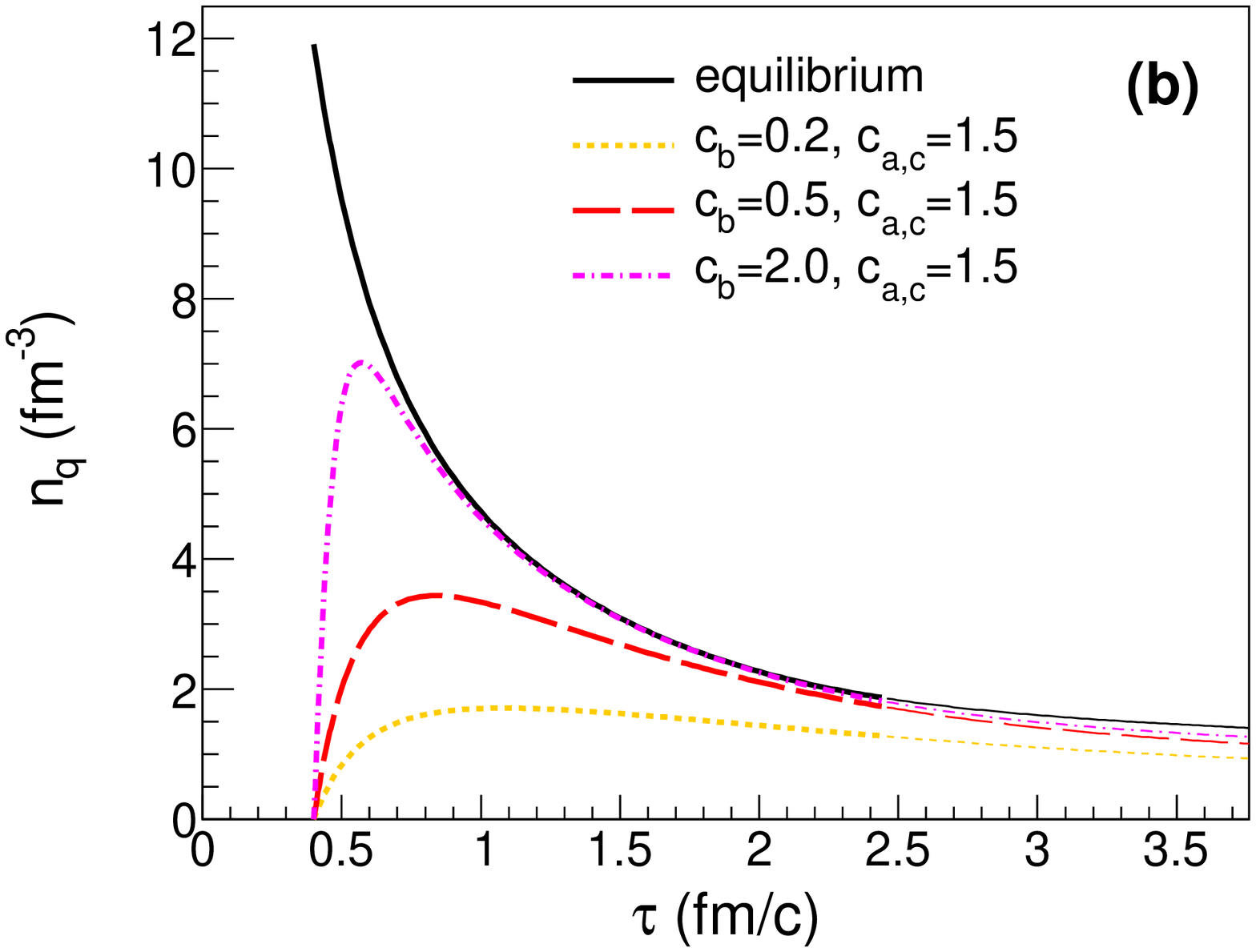} \hspace{1pc}%
\begin{minipage}[b]{9pc}\caption{\label{label} (a) Thermal photon $v_2 (p_T)$ and (b) quark number density $n_q (t)$ for different quark chemical equilibration rates. \label{fig:2}}
\end{minipage}
\end{figure}

Differential elliptic flow of thermal photons $v_2 (p_T)$ and time evolution of the quark number density $n_q(t)$ in and out of chemical equilibrium are shown in Fig.~\ref{fig:2}. $v_2 (p_T)$ is found to become visibly larger for slower quark production because the contribution of hadronic photons becomes effectively large. The quark chemical reaction rate parameter is varied as $c_b = 0.2, 0.5$ and $2.0$ where $c_a = c_c = 1.5$ are fixed. The equilibrium result is recovered in the large $c_b$ limit. Estimated chemical relaxation times for the chosen parameters are $\tau_\mathrm{ch} \sim 1/c_b T \sim 2.0, 0.5$ and $0.2$ fm/$c$ for the average temperature $T\sim 0.2$ GeV, which are consistent with the numerical results. It should be noted that the parton density does not converge to that in equilibrium as defined by the parton picture near and below the crossover (indicated by thin lines in the figure) because the information of the hadronic phase is embedded in the EoS.

\begin{figure}[tbh]
\includegraphics[width=13.5pc]{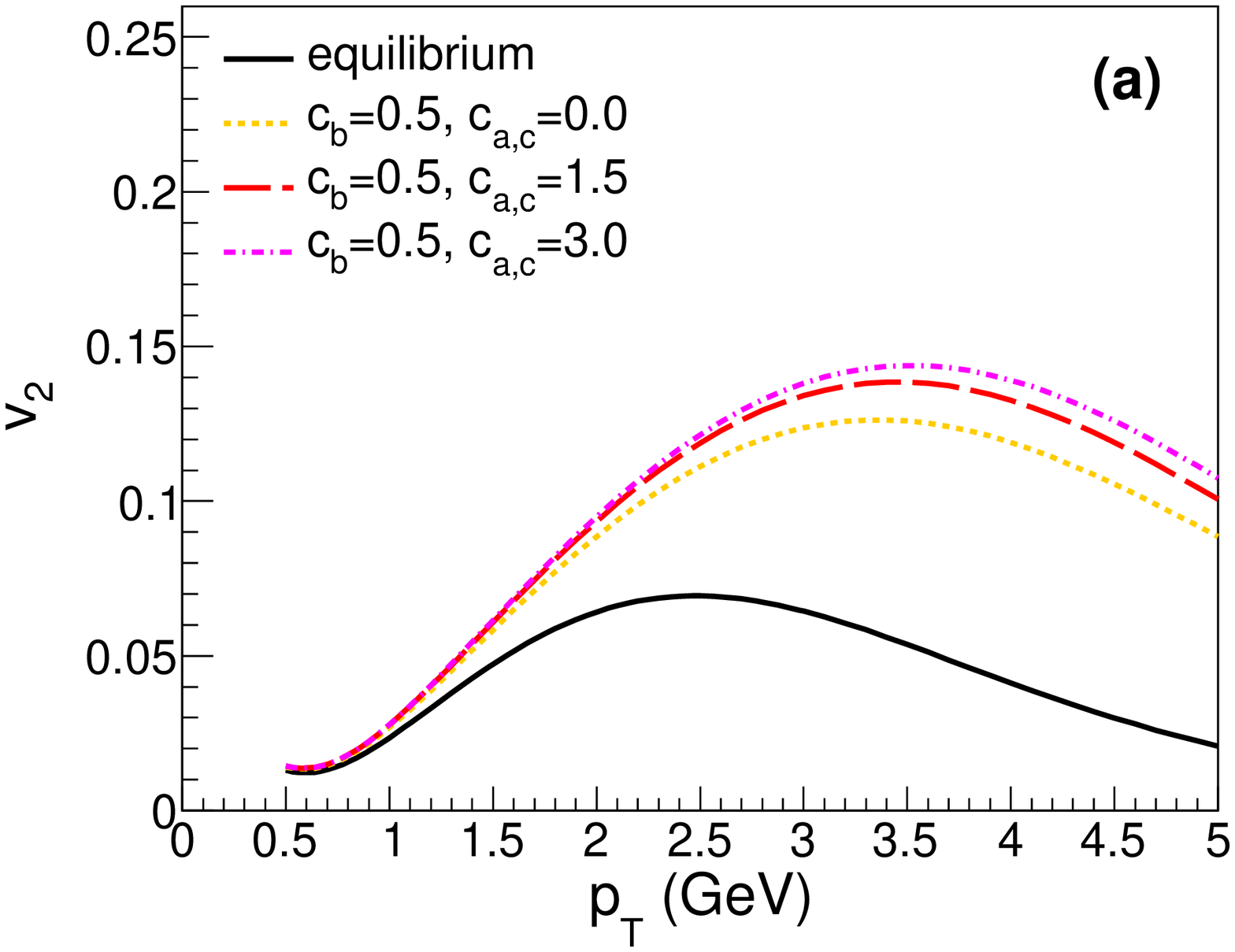} 
\includegraphics[width=13.5pc]{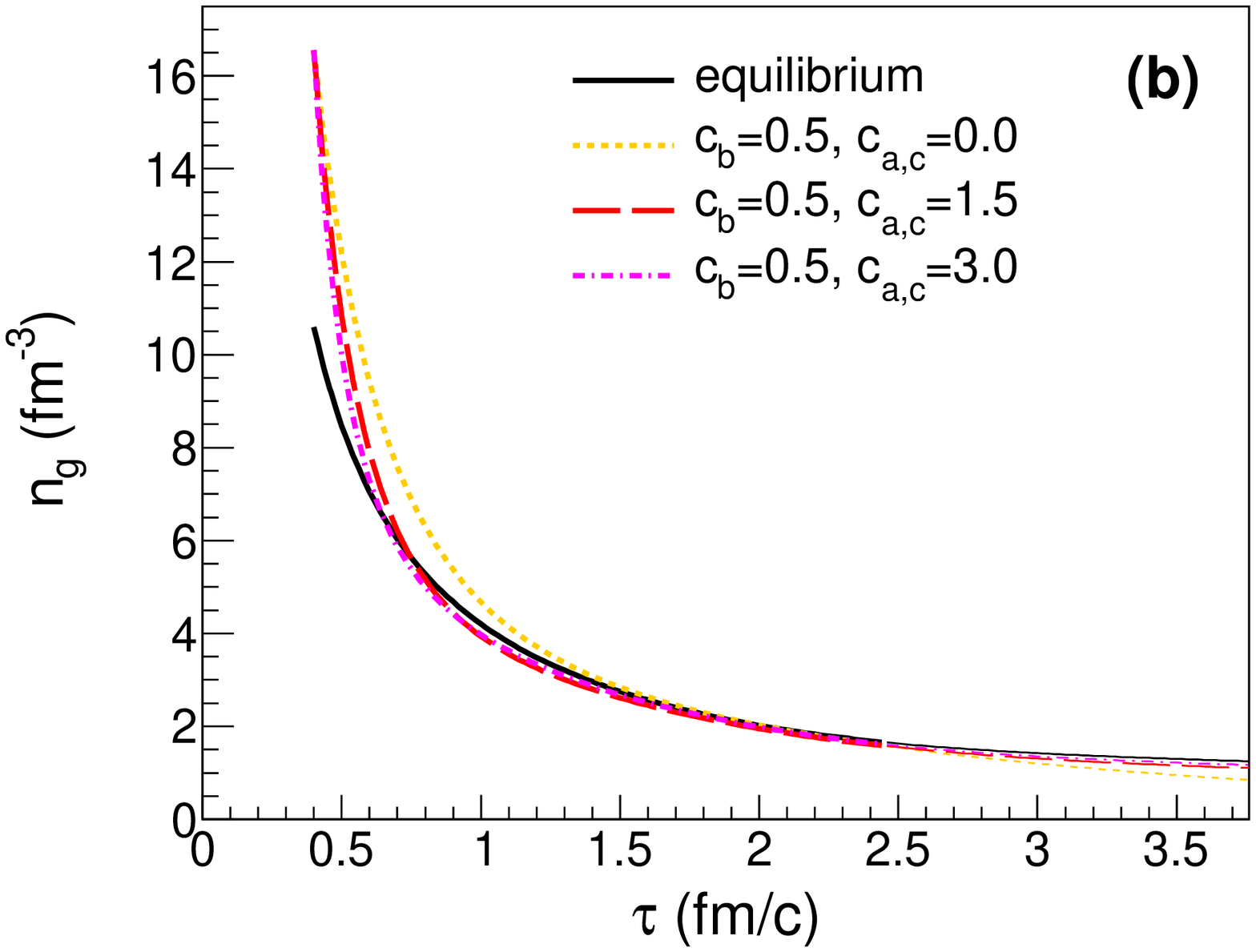} \hspace{1pc}%
\begin{minipage}[b]{9pc}\caption{\label{label} (a) Thermal photon $v_2 (p_T)$ and (b) quark number density $n_g (t)$ for different gluon chemical equilibration rates. \label{fig:3}}
\end{minipage}
\end{figure}

Figure~\ref{fig:3} shows the numerical results for several different choices of the chemical equilibration rates for gluon splitting/merging and gluon emission from quarks. The reaction rate parameters are varied from $c_a = c_c = 0.0, 1.5$ and $3.0$ for the fixed $c_b = 0.5$. One can see that $v_2 (p_T)$ becomes smaller for slower quark-number conserving chemical equilibration, though the dependences on $c_a$ and $c_c$ are weaker than that on $c_b$. This is because gluons are initially overpopulated and the recombination processes become dominant for gluons. The quark number density is indirectly suppressed by faster gluon recombination at early times. 

A source of overestimation is the fact that vanishing quark number is assumed in the initial condition, while initial sea quarks and pre-thermal chemical equilibration will prepare finite number of quarks at the beginning of the hydrodynamic stage. This is effectively taken into account by varying the reaction rates. Also dynamical effects on the EoS itself is important \cite{Gelis:2004ep}, as thermal photons are typically sensitive to it. 

\section{Conclusion and outlook}
I estimated the effects of quark chemical equilibration in the hydrodynamic stage on thermal photon elliptic flow. A newly-developed (2+1)-dimensional ideal hydrodynamic model is coupled to the rate equations for parton number densities for the QGP phase. Thermal photon $v_2$ is found to become visibly larger when the quark chemical equilibration is slow because of the initial suppression of the photons with small anisotropy. It should be noted that the medium can initially be gluon-rich due to the fact that the colliding nuclei are described as the color glass condensate. This can partially explain the excessive photon $v_2$ recently observed in the collider experiments. Late gluon chemical equilibration slightly reduces $v_2$ but the dominant process for the modification of the quantity is quark pair production and annihilation. The results imply that chemical composition of a QCD medium and subsequent equilibration processes are important aspects in heavy-ion physics. 

Future prospects include the estimation of prompt photons, which might wash some of the enhancement effect out of direct photon $v_2$. Also chemical equilibration rates should be constrained from the viewpoint of microscopic theory in the same way viscosity and equation of state are done. Modification of transport coefficients due to off-equilibrium chemical composition would be worth-investigating. Additional $v_2$ enhancement mechanism \cite{Monnai:2014taa} may have to be introduced as excessive initial suppression tends to reduce photon particle spectra.

\ack
The work is inspired by fruitful discussion with M\"{u}ller B. The author would like to thank for valuable comments by Akiba Y and McLerran L on the paper. The work is supported by RIKEN Special Postdoctoral Researcher program.

\end{document}